\documentclass[preprint,12pt]{elsarticle}
\usepackage{amsmath,amssymb}
\usepackage{graphicx}

\begin{document}

\title{Fermi-surface reconstruction involving two van Hove singularities across the antiferromagnetic transition in BaFe$_2$As$_2$}

\author[a,b]{Y. Nakashima\corref{cor}}\ead{yosuke-nakashi-march@hiroshima-u.ac.jp}
\author[a,b]{A. Ino\corref{cor}}\ead{ino@hiroshima-u.ac.jp}
\author[a,b]{S. Nagato}
\author[c]{H. Anzai}
\author[c]{H. Iwasawa}
\author[a,b]{Y.~Utsumi}
\author[b,c]{H.~Sato}
\author[c]{M. Arita}
\author[c]{H.~Namatame}
\author[a,c]{M. Taniguchi}
\author[d]{T.~Oguchi}
\author[b,e]{Y.~Aiura}
\author[e]{I.~Hase}
\author[b,e]{K. Kihou}
\author[b,e]{C. H. Lee}
\author[b,e]{A.~Iyo}
\author[b,e]{H.~Eisaki}
\cortext[cor]{Corresponding author}

\address[a]{Graduate School of Science, Hiroshima University, Higashi-Hiroshima 739-8526, Japan}
\address[b]{JST, Transformative Research-Project on Iron Pnictides (TRIP), Chiyoda, Tokyo 102-0075, Japan}
\address[c]{Hiroshima Synchrotron Radiation Center, Hiroshima University, Higashi-Hiroshima 739-0046, Japan}
\address[d]{Institute of Scientific and Industrial Research, Osaka University, Ibaraki, Osaka 567-0047, Japan}
\address[e]{National Institute of Advanced Industrial Science and Technology, Tsukuba 305-8568, Japan}

\date{\today}

\begin{abstract}
We report an angle-resolved photoemission study of BaFe$_2$As$_2$, a parent compound of iron-based superconductors.  Low-energy tunable excitation photons have allowed the first observation of three-dimensional saddle-point dispersion at the $Z$ point, as well as the $\Gamma$ point.  This resolves a discrepancy in Fermi-surface studies.  We found that these van Hove singularities are not only involved in three-dimensional Fermi-surface reconstruction with antiferromagnetic ordering but also renormalized and predominant at low energies.  Hence, the Fermi-surface instability due to these singularities would also underlie the associated superconductors.
\end{abstract}


\maketitle

\section{Introduction}
\label{Introduction}
For the newly-discovered iron-based superconductors \cite{KamiharaY2008JACS}, essential ingredients are considered to be inherited from their antiferromagnetic (AF) parent compounds.  Carrier doping, pnictogen substitution, and pressure application can all give rise to a superconducting (SC) phase out of a collinear AF metallic phase.  Such phase transitions commonly evolve from Fermi-surface instabilities.  For example, the nesting of the Fermi surfaces induces some density-wave order associated with gap opening.  Thus the nesting condition for the AF order has been extensively examined \cite{ZabolotnyyVB2009Nature,LiuC2009PRL,MalaebW2009JPSJ,ThirupathaiahS2010PRB}.  Another typical instability occurs near a van Hove singularity (vHS)---a sharp edge in the density of states (DOS), arising from a stationary point of band dispersion \cite{HalbothCJ2000PRL,YamaseH2000JPSJ,ZhaiH2009PRB}.  There, a slight energy shift of the vHS is enough to reduce the DOS at the Fermi level ($E_\mathrm{F}$) for stabilization in association with a topological change of the Fermi surface.  The excitations responsible for the Fermi-surface instabilities, such as spin fluctuations near the AF phase, are the candidates for the SC pairing glue.  Therefore, the Fermi surface of the parent compound serves as a ground for exploring the driving force of the transitions to the AF and SC phases.

However, despite intensive studies, the Fermi surface of the AF phase of the 122-type iron arsenides has been controversial.  The most distinctive discrepancy is about perpendicular-momentum ($k_z$) dependence.  Many angle-resolved photoemission (ARPES) studies using excitation photons of conventional energies, $h\nu=25-105$ eV, have reported that the innermost Fermi surface along the $k_z$ axis is of a large volume, and that its cross-section area monotonically increases in going from the $k_z=0$ plane to the $k_z=2\pi/c$ plane of Brillouin-zone (BZ) boundary \cite{VilmercatiP2009PRB,MalaebW2009JPSJ,LiuC2009PRL,YiM2009PRB,KondoT2010PRB,ThirupathaiahS2010PRB,WangQ2010arxiv}.  This is qualitatively similar to that given by first-principles calculations for the non-magnetic (NM) state \cite{XuG2008EPL,MaF2010FPC}, implying that the $k_z$ dependence is not so affected by the AF order.  On the other hand, quantum-oscillation studies have indicated that the Fermi surface along the $k_z$ axis is of a small volume and closed before reaching the $k_z=2\pi/c$ plane \cite{SebastianSE2008JPCM,AnalytisJG2009PRB,TerashimaT2011PRL}.  Their results are almost consistent with first-principles calculations for the collinear AF state, indicating a three-dimensional (3D) Fermi-surface reconstruction due to the AF order \cite{MaF2010FPC,SebastianSE2008JPCM,AnalytisJG2009PRB,TerashimaT2011PRL}.  This picture has also been argued from a laser-based low-energy ARPES study using $h\nu=7$ eV, although the data are confined to a certain $k_z$ plane \cite{ShimojimaT2010PRL}.  Between the contrasting ARPES reports, a considerable difference is in the excitation-photon energy $h\nu$.  As $h\nu$ decreases from 50 to 13 eV, the typical probing depth increases from $\sim$5 to $\sim$20 {\AA} \cite{SeahDench1979} in addition to the improvement in momentum resolution \cite{YamasakiT2007PRB}.  Therefore, a $k_z$-dependent low-energy ARPES study is required for further clues to the bulk Fermi surface.

In this paper, we report the 3D shapes of the Fermi surface of BaFe$_2$As$_2$, using ARPES spectra taken by sweeping $h\nu$ in a low-energy range from 6.5 to 30 eV.  On the basis of the first observation of non-monotonic dispersion along the $k_z$ axis, we investigate two saddle-point dispersions from the aspect of their behaviors on the AF transition and their predominance in low-energy DOS, and then briefly discuss a Fermi-surface instability in possible relation to the superconductivity.

\section{Experimental}
\label{Experimental}
\begin{figure}
\includegraphics[width=\columnwidth,bb=0 0 646 510]{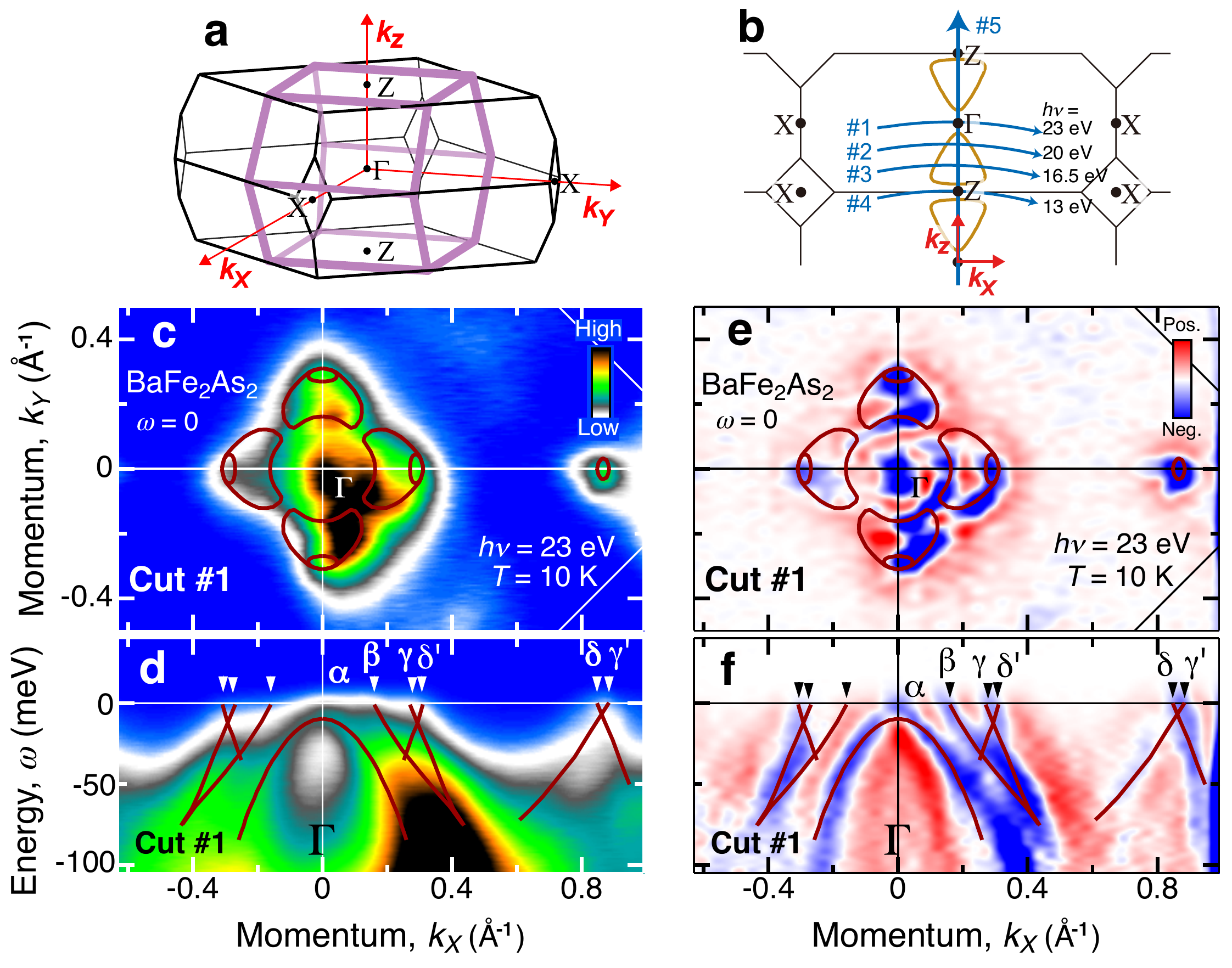}\vspace{-6pt}
\caption{(a) Notation of point labels and axes. Thick purple and thin black lines denote the Brillouin zones (BZs) with and without the AF order, respectively.  (b) Cutting paths \#1-\#5, depicted in the $k_Y=0$ plane.  (c),(d) Fermi-surface mapping and energy-momentum distribution of ARPES intensity in the AF state at $T=10$ K, taken along cut \#1 through $\Gamma$.  (e),(f) Same as (c) and (d), respectively, but filtered by $k_X$-$k_Y$ Laplacian.  The red curves in (c)-(f) are based on the Laplacian-filtered ARPES-intensity maps.}
\label{inplane}
\end{figure}

High-quality single crystals of BaFe$_2$As$_2$ were grown by a flux method, as described in Ref.~\cite{NakajimaM2010PRB}. ARPES measurements were performed at a helical undulator beamline, BL-9A of the Hiroshima Synchrotron Radiation Center (HRSC), using a Scienta R4000 electron analyzer.  The total energy and momentum resolutions were typically 12 meV and 0.01 {\AA}$^{-1}$, respectively, for $h\nu=13$ eV.  The samples were cleaved \textit{in situ} and kept under ultra-high vacuum better than $4\times 10^{-11}$ Torr during the measurements.  First-principles density-functional calculations were performed using an all-electron full-potential linear augmented-plane-wave method.  Although the BZ of BaFe$_2$As$_2$ should be folded back under the collinear AF order below $T_\mathrm{N}=134$ K, we use the labels of the unfolded tetragonal BZ throughout this paper as shown in Fig.~\ref{inplane}(a), and define the $k_X$ axis as the $\Gamma$-$X$ direction.  The in-plane cuts \#1-\#4, which intersect the $k_z$ axis at $k_z c/4\pi=3.00$, 2.86, 2.69 and 2.50 as shown in Fig.~\ref{inplane}(b), were taken with $h\nu=23$, 20, 16.5 and 13 eV, respectively, which were given by a free-electron final-state approximation and an inner potential of 13.6 eV determined from the perpendicular cut \#5 presented in Fig.~\ref{3d}(c5). 

\section{Results}
\label{Results}
In the AF state at $T=10$ K, the BZ folding is easily seen in the in-plane cuts.  The spectral-intensity maps are directly shown in Figs.~\ref{inplane}(c) and (d), and those applied with a two-dimensional $k_X$-$k_Y$ Laplacian filter, $\nabla_\mathrm{k}^2 = (\partial/\partial k_X)^2 + (\partial/\partial k_Y)^2$, are shown in Figs.~\ref{inplane}(e) and (f).  The valley curves of the Laplacian plots are denoted by the red curves in Figs.~\ref{inplane}(c)-(f).  As shown in Figs.~\ref{inplane}(d) and (f), three hole-pocket-like bands and an electron-pocket-like band are observed around $\Gamma$, and labeled $\alpha$, $\beta$, $\gamma$ and $\delta^\prime$, respectively.  The dispersions labeled $\gamma^\prime$ and $\delta$ near $X$ are almost identical to those of $\gamma$ and $\delta^\prime$, respectively, when shifted in momentum by a vector connecting $\Gamma$ and $X$, exhibiting the BZ folding due to the AF order.  Since the $\delta^\prime$ and $\gamma^\prime$ bands are unexpected from the NM calculations, they are regarded as replica bands \cite{XuG2008EPL,MaF2010FPC}.  Figures~\ref{inplane}(c) and (e) show complicated flower-like Fermi surface, indicating that the band structure is further reconstructed by hybridization, as recognized previously \cite{ShimojimaT2010PRL,LiuG2009PRB,YiM2009PRB}.  Here, the fourfold symmetry is due to the coexistence of structural twin domains \cite{TanatarMA2009PRB}.

\begin{figure}
\begin{center}
\includegraphics[width=0.75\columnwidth,bb=0 0 775 1213]{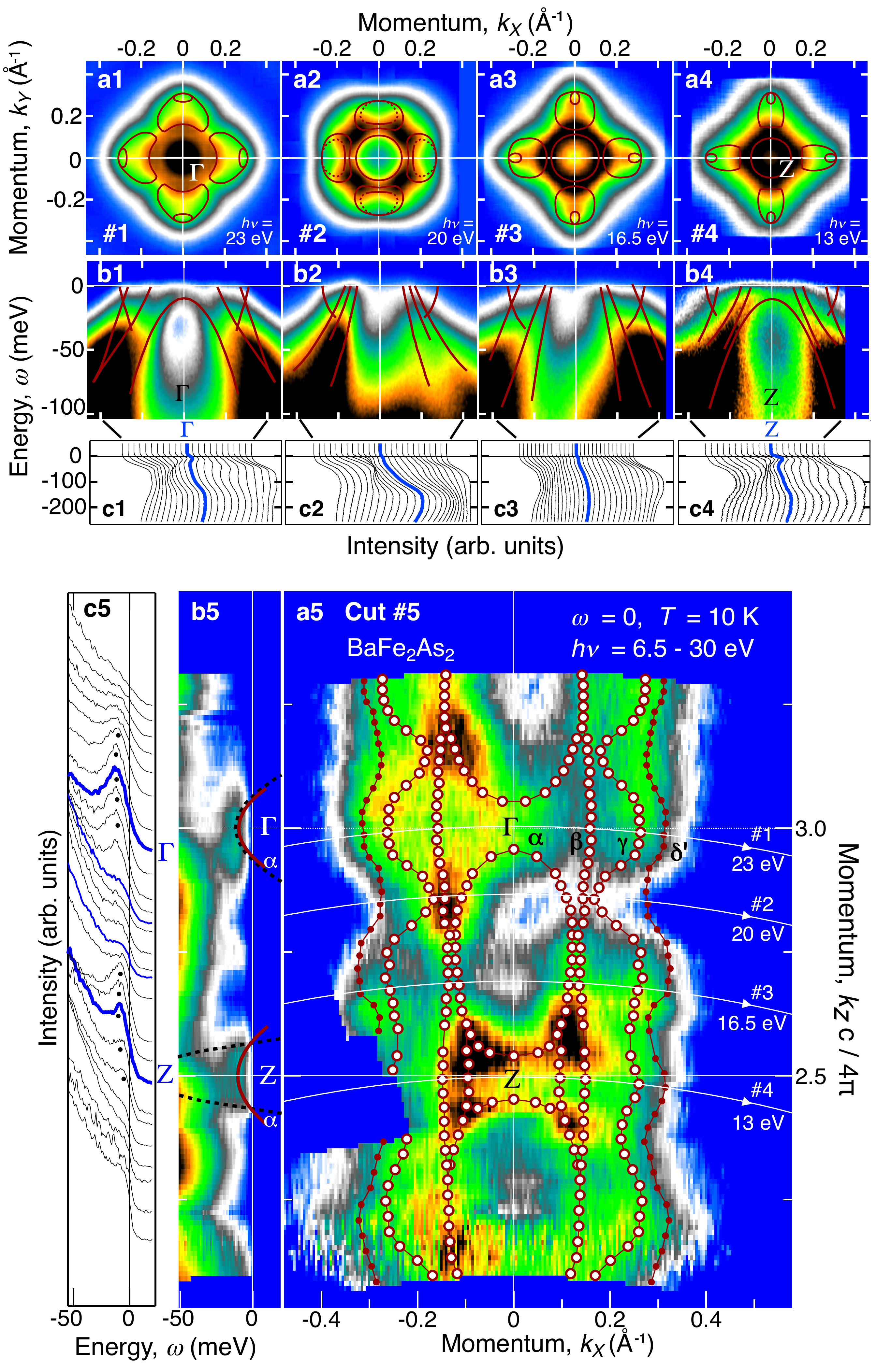}\vspace{-6pt}
\caption{(a1)-(a4) Fermi-surface mappings along the in-plane cuts \#1-\#4 indicated in Fig.~\ref{inplane}(b).  The intensity maps were symmetrized about the $k_X=0$ and $k_X=k_Y$ lines.  (a5) Fermi-surface mapping along the perpendicular cut \#5 ($k_Y=0$ plane).  (b1)-(b5) Energy-momentum distributions along the cuts \#1-\#5.  (c1)-(c5) Energy distribution curves on the cuts \#1-\#5.  All the red curves in (a1)-(a4) and (b1)-(b5) are based on the Laplacian-filtered ARPES-intensity maps.  As for (b5), the dispersion predicted by the first-principle calculation is also shown by black dotted curves for comparison.}
\label{3d}
\end{center}
\end{figure}

Comparing the in-plane cuts \#1-\#4, one finds a dramatic $k_z$ dependence in the AF state.  In particular, it should be noted that the innermost band $\alpha$ shows non-monotonic $k_z$ dispersion.  As highlighted with blue curves in Figs.~\ref{3d}(c1)-(c4), a spectral peak remains below $E_\mathrm{F}$ along the cuts \#1 and \#4 passing through $\Gamma$ and $Z$, respectively, while it crosses $E_\mathrm{F}$ and then disappears along cuts \#2 and \#3.  As shown in Figs.~\ref{3d}(b1)-(b4), the innermost hole pocket $\alpha$ has its top distinctly below $E_\mathrm{F}$ for cuts \#1 and \#4, while it is across $E_\mathrm{F}$ for cuts \#2 and \#3.  In Figs.~\ref{3d}(a1)-(a4), the intensity distribution is seen to peak at the center for cuts \#1 and \#4, while its peak is ring-shaped around the center for cuts \#2 and \#3.  

The perpendicular cut \#5 complements the 3D perspective.  We identified $\Gamma$ and $Z$ as the 3D saddle points of the $\alpha$ band, because Figs.~\ref{3d}(b5) and (c5) show that the bottoms of the $k_z$ dispersion are at $\Gamma$ and $Z$, where the top of the in-plane dispersion is located.  The observation of clear Fermi-surface crossings in Figs.~\ref{3d}(b5) and (c5) provides evidence that these saddle points separate the innermost hole pockets $\alpha$.  The mapping along the $k_z$ axis in Fig.~\ref{3d}(a5) shows that the $\alpha$ pocket is tear-shaped and indeed of a small volume, and additionaly that some features are more dispersive in $k_z$ than those observed previously \cite{VilmercatiP2009PRB,MalaebW2009JPSJ,LiuC2009PRL,YiM2009PRB,KondoT2010PRB,ThirupathaiahS2010PRB,WangQ2010arxiv}.  Not only the $\alpha$ pocket, but also the $\gamma$ segment of a Fermi surface is tightly curved around $k_z\simeq 12\pi/c$.  Nevertheless, at a majority of $k_z$ values, the $\gamma$ and $\delta^\prime$ segments are close to each other, suggesting that the nesting is still realized between them.

\begin{figure} 
\includegraphics[width=\columnwidth,bb=0 0 559 360]{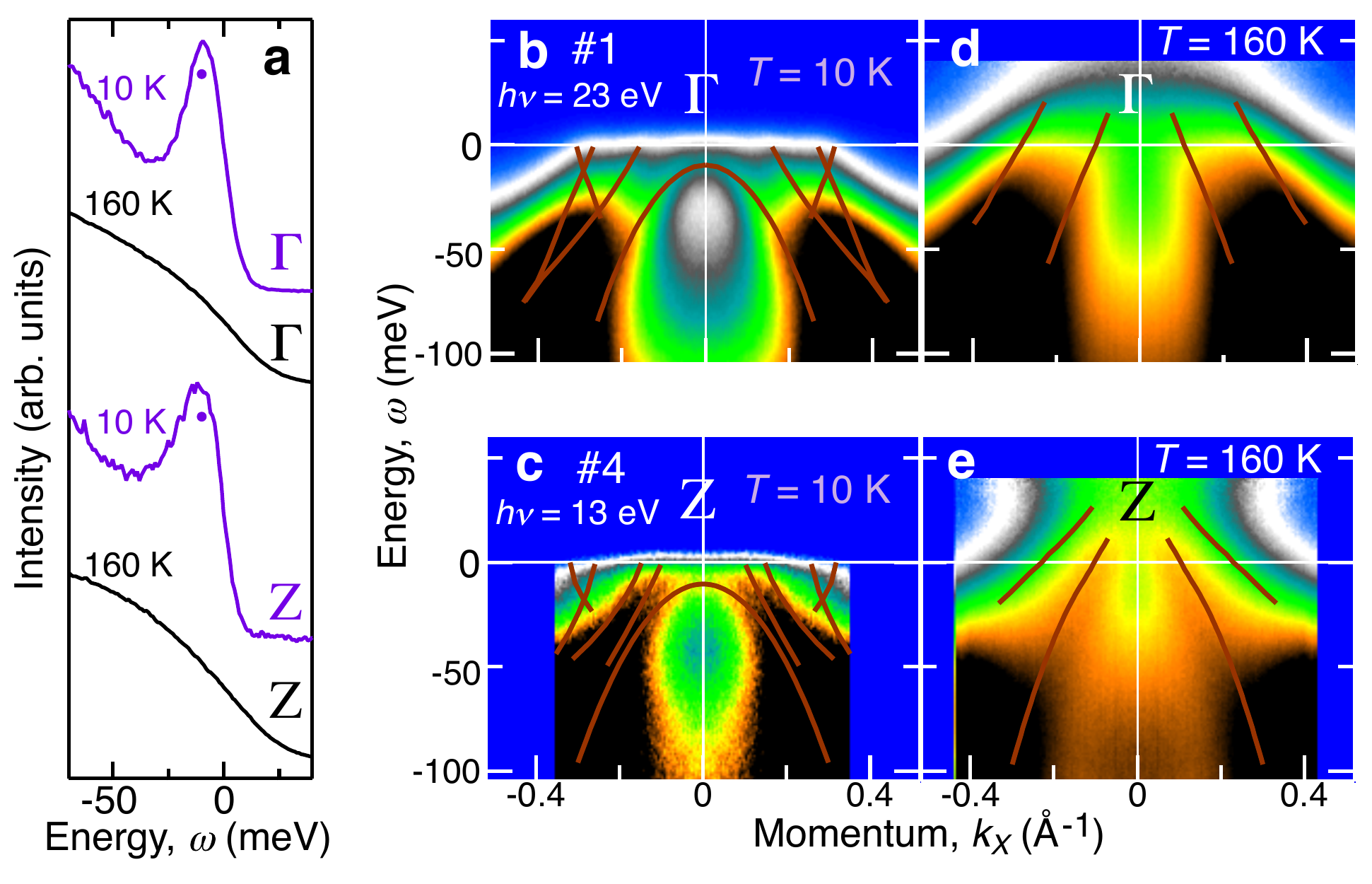}\vspace{-6pt}
\caption{Spectral change across the AF transition temperature $T_\mathrm{N}=134$ K.  (a) Energy distribution curves at $\Gamma$ and $Z$.  (b),(c) Energy-momentum distributions at $T=10$ K, taken along the cuts \#1 and \#4 through $\Gamma$ and $Z$, respectively, and symmetrized about the $k_X=0$ line.  (d),(e) Same as (b) and (c), respectively, but at $T=160$ K and divided by the Fermi-Dirac distribution function.  All the red curves are based on the Laplacian-filtered ARPES-intensity maps.}
\label{temperature}
\end{figure}

\begin{figure}
\includegraphics[width=\columnwidth,bb=0 0 839 561]{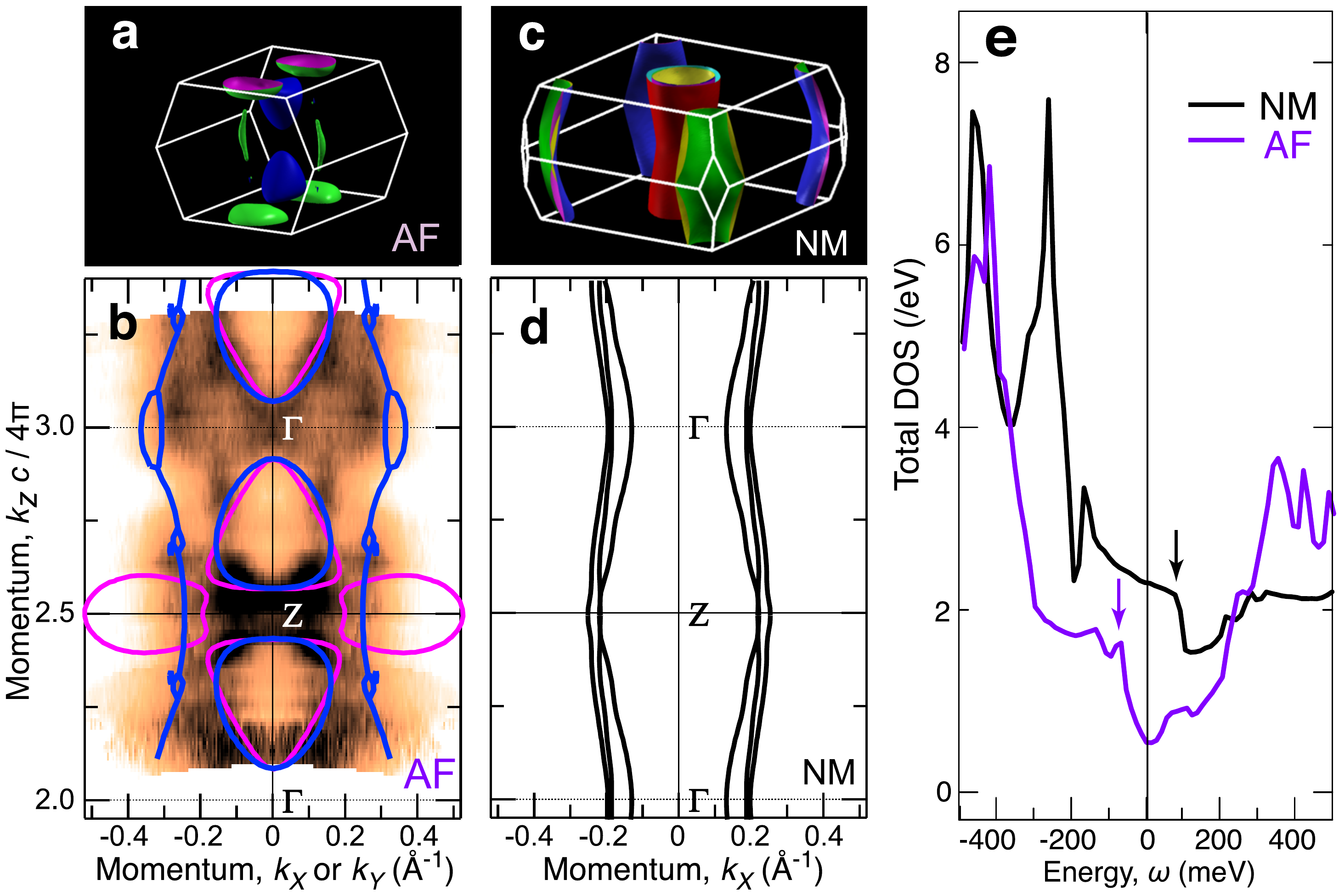}\vspace{-6pt}
\caption{Results of first-principles calculation.  (a),(b) Perspective and cross-sectional views of the Fermi surface for the collinear antiferromagnetic (AF) state.  The calculations along the twin axes, $k_X$ and $k_Y$, are denoted by the blue and purple curves in (b), respectively, and compared with the symmetrized ARPES-intensity map from Fig.~2(a5).  (c),(d) Same as (a) and (b), respectively, but for the non-magnetic (NM) state.  (e) Total DOS for the NM (black) and AF (violet) states with arrows indicating the vHSs near $E_\mathrm{F}$.}
\label{calc}
\end{figure}

Figure~\ref{temperature} shows that the 3D-saddle-point energies are involved in a phase transition.  At $T=10$ K, the spectral peaks at $\Gamma$ and $Z$ are distinct, and their hole-pocket-like in-plane dispersions are observed below $E_\mathrm{F}$, as summarized in Figs.~\ref{temperature}(a)-(c).  When the temperature is raised to 160 K across $T_\mathrm{N}$, the peaks disappear from $\Gamma$ and $Z$, and the in-plane dispersions cross $E_\mathrm{F}$ with no spectral weight staying below $E_\mathrm{F}$, as shown in Figs.~\ref{temperature}(a), (d) and (e).  These drastic spectral changes provide evidence that the AF phase transition is accompanied by a topological change of the innermost Fermi surface $\alpha$ from cylindrical to tear shapes with respect to the $\Gamma$ and $Z$ saddle points. 

The involvement of these 3D saddle points is also indicated by our first-principles calculation.  When the AF order is incorporated into the calculation, the innermost hole pocket $\alpha$ topologically changes across the $\Gamma$ and $Z$ saddle points, as shown in Figs.~\ref{calc}(a)-(d), and the corresponding shifts of the vHSs across $E_\mathrm{F}$ result in a considerable reduction of the DOS at $E_\mathrm{F}$, as shown in Fig.~\ref{calc}(e).

\section{Discussion}
\label{Discussion}
Our result shows some qualitative differences from the conventional ARPES data on the AF state.  Notably, the non-monotonic $k_z$ dispersion in Fig.~\ref{3d} contrasts with the previously-reported simple monotonic $k_z$ dispersion \cite{VilmercatiP2009PRB,MalaebW2009JPSJ,LiuC2009PRL,YiM2009PRB,KondoT2010PRB,ThirupathaiahS2010PRB,WangQ2010arxiv}, which has been interpreted along the calculations without the AF order.  Rather, the present result shares some essential features with the results of the bulk-sensitive quantum-oscillation studies \cite{SebastianSE2008JPCM,AnalytisJG2009PRB,TerashimaT2011PRL} and likewise with the calculation for the collinear AF state.  In particular, the small tear-shaped $\alpha$ pocket brought out in Fig.~\ref{3d}(c5) is well reproduced by the calculation in Figs.~\ref{calc}(a) and (b).  Although Fig.~\ref{3d}(c5) shows apparent coexistence of two $\alpha$ sheets parallel and perpendicular to the $k_z$ axis around $Z$, these distinct sheets are indeed predicted for the coexisting twin axes, as seen in Figs.~\ref{calc}(a) and (b).  The outer band $\gamma$ seen at $k_X\simeq 0.26$ {\AA}$^{-1}$ in Fig.~\ref{3d}(c5) is reminiscent of the cigar-shaped pockets in the calculation.  Our observation of the more detailed $k_z$ dispersions than before is possibly due to the low-energy excitation photons, because they allow the photoemission from multiple Fe-As layers beneath the sample surface, and the well-defined photoemission final states may contribute to the $k_z$ resolution.  Thus, the difference between the ARPES results may imply that the magnetic and electronic states are affected near the surface.

Nevertheless, some experimental features are not reproduced by the calculation.  For example, the petal-shaped electron pockets observed in all the in-plane cuts \#1-\#4 [Figs.~\ref{3d}(a1)-(a4)] are absent in the calculation except for cut \#4 [Fig.~\ref{calc}(a)].  Although some residual signals from the sample surface cannot be excluded, the discrepancies are probably associated with the fact that the calculated magnetic moment, $1.58\mu_\mathrm{B}$ per Fe site, is substantially larger than the experimental value, $0.93\mu_\mathrm{B}$, from neutron diffraction~\cite{HuangQ2008PRL}.  Accordingly, it is reasonable to assume that the magnitude of the AF gap and the extent of Fermi-surface reconstruction are, in reality, not as much as the calculation predicts.

\begin{table}
\caption{Energy, $\omega_\mathrm{SP}$, and effective masses, $m_\parallel$ and $m_\perp$, of $\Gamma$ and $Z$ saddle points, determined by the ARPES experiment and the first-principles calculation for BaFe$_2$As$_2$.  The effective masses are deduced from the dispersion curvatures along the in-plane and perpendicular directions, and given in unit of free-electron mass, $m_\mathrm{e}$.}
\label{vHSs}
\begin{center}
\begin{tabular}{cp{8pt}ccp{6pt}cc}
\hline \hline
		&& $\Gamma$(exp.) & $\Gamma$(calc.) && $Z$(exp.) & $Z$(calc.) 	\\
\hline
$\omega_\mathrm{SP}$ (meV)  && $\!\!-10(2)$  & $\!\!-12$ && $\!\!-10(2)$ & $\!\!-82$ 	\\
$m_\perp/m_\mathrm{e}$ && $1.2(8)$ & $1.3$ && $1.7(9)$ & $0.21$ 	\\ 
$-m_\parallel/m_\mathrm{e}$ && $3.5(6)$ & $\;1.9$\footnotemark[1] && $4.0(6)$ & $2.9$\footnotemark[1] 		\\
&&& $\;\;0.68$\footnotemark[2] &&& $\!>$20\footnotemark[2] \\
\hline \hline
\end{tabular}
\end{center}
\footnotetext[1]{$k_X$ direction, depicted in Fig.~\ref{inplane}(a).}
\footnotetext[2]{$k_Y$ direction, likewise.}
\end{table}

Next, we focus on the experimental characteristics of 3D-saddle-point singularities.  Qualitatively, the present result is consistent with the calculation, in regard to the saddle-point shifts across $E_\mathrm{F}$ at $\Gamma$ and $Z$ as a result of the AF transition.  Quantitatively, the energy renormalizations with respect to the first-principles calculation were identified, as listed in Table~\ref{vHSs}.  Around $\Gamma$, the in-plane dispersion curvature is flattened by a factor of 2 or 5.  Around $Z$, both the $k_z$-dispersion curvature and the saddle-point energy are strongly renormalized by a factor of $\sim$8, which is larger than the $\alpha$-pocket average, $\sim$3, from quantum oscillations \cite{TerashimaT2011PRL}, as shown by the black dotted and red solid curves in Fig.~\ref{3d}(b5).  Their roles in phase-transition physics may be inferred from the DOS.  The calculation in Fig.~\ref{calc}(e) shows that the DOS sharply drops near $E_\mathrm{F}$ to $\sim$1/3 for the AF state.  Experimentally, the above strong renormalizations make this singularity even higher, sharper and closer to $E_\mathrm{F}$.  The shifts of such vHSs across $E_\mathrm{F}$ should have substantial impacts on the total energy, suggesting their contributions to the stabilization of the ordered phase.

These findings provide a new perspective on the fluctuations underlying the associated superconductors.  The nesting instability of the outer Fermi surfaces has been noted previously\cite{ZabolotnyyVB2009Nature,LiuC2009PRL,MalaebW2009JPSJ,ThirupathaiahS2010PRB,KondoT2010PRB,WangQ2010arxiv}, and is compatible with our data at the majority of $k_z$.  However, the absence of nesting has been reported for another iron-based superconductor LiFeAs \cite{BorisenkoSV2010PRL}.  Alternatively, the saddle-point instabilities of the innermost Fermi surface can also be expected near the AF phase from the above findings on the vHSs, because the innermost Fermi surface itself has a strong tendency to change from cylindrical to tear-shaped at the cost of modification of the other electronic states.  In general, a Fermi-surface instability is specifically enhanced by involving a vHS.  This has long been argued as a possible scenario for the high-temperature superconductivity in cuprates, and also for the spontaneous rotational-symmetry breaking into the electron-nematic state, as known for Pomeranchuk instability \cite{HalbothCJ2000PRL,YamaseH2000JPSJ,ZhaiH2009PRB,FradkinE2010Science,ChuangTM2010Science}.  The vHSs dominating the AF transition could also be the key points in the energetics of the SC transition.

\section{Conclusion}
\label{Conclusion}
In conclusion, we observed the totally-reconstructed 3D Fermi surface alongside the $\Gamma$ and $Z$ saddle points below $E_F$ in the AF state of BaFe$_2$As$_2$, resolving a major discrepancy between the ARPES results and the bulk-sensitive information from quantum oscillations and first-principles calculations.  We revealed that the 3D-saddle-point singularities near the innermost Fermi surface are enhanced by energy renormalization and involved in the AF transition.  The possible role of the saddle-point instabilities in the iron-based superconductors needs to be addressed in depth by future studies.

\section*{Acknowledgement}
\label{Ack}
We acknowledge discussions with K.~Shimada and A.~Kimura, and technical support from J.~Jiang and H.~Hayashi.  This work was supported by KAKENHI (20740199).  The ARPES experiments were done under the approval of HSRC (Proposal No.~09-A-1 and 10-A-14).

\end{document}